\pdfoutput=1
\documentclass[11pt,a4paper]{article}
\usepackage{amsmath,amssymb}
\usepackage{braket}
\usepackage{url}
\usepackage{mathtools}
\usepackage{amsmath}              
\usepackage{graphics,graphicx,epsfig,ulem} 
\usepackage{subfigure,subfig}
\usepackage{feynmp}
\usepackage{slashed}

\renewcommand\[{\left[}
\renewcommand\]{\right]}

\def\beq{\begin{equation}}
\def\eeq{\end{equation}}
\def\[{\begin{equation}}
\def\]{\end{equation}}

\begin{document}
\numberwithin{equation}{section}

\title{
\vspace{2.5cm}
\Large{\textbf{Multiparticle Higgs and Vector Boson Amplitudes at Threshold}}}

\author{Valentin V. Khoze\footnote{valya.khoze@durham.ac.uk}\\[4ex]
  \small{\it Institute for Particle Physics Phenomenology, Department of Physics} \\
  \small{\it Durham University, Durham DH1 3LE, United Kingdom}\\[0.8ex]
}

\date{}
\maketitle

\begin{abstract}
  \noindent

\end{abstract}
In a spontaneously broken gauge theory we consider (sub)-processes in which one virtual intermediate state (it can be a Higgs or a gauge field)
produces many on-shell Higgses and massive vector bosons. In the kinematic regime where all final states are produced on their mass threshold, we show
how to compute iteratively all tree-level amplitudes  ${\cal A}_{1\to n+m}$  involving an arbitrary number $n$ of Higgs bosons and $m$ of longitudinal vector bosons
in the final state, and list the amplitudes coefficients for up to $n=32$ and $m=32$. We find that these amplitudes exhibit factorial growth not only in the number of scalar
fields, but also in the number of longitudinal gauge fields, ${\cal A}_{1\to n+m} \sim n! \,m!$. This growth is not expected to disappear at loop-level in the fixed-order perturbation 
theory. We conclude that at energies accessible at the next generation of hadron colliders, such as the 50-100 TeV FCC, where $\sqrt{\hat{s}}$ is sufficient to produce $\gg 1/\alpha_W$ of $W,Z$ and $H$, 
perturbation theory breaks down when applied to the multiparticle electroweak production, at least near the kinematic multiparticle mass threshold where the electroweak 
gauge-Higgs sector becomes strongly coupled.
\bigskip
\thispagestyle{empty}
\setcounter{page}{0}

\newpage

\section{Introduction}\label{sec:intro}

At sufficiently high energies it becomes kinematically possible to produce a high multiplicity final state with $n\sim 1/\alpha$ particles
in the weakly interacting theory. In this case the well-known problem of the factorial divergencies \cite{Dyson} of large orders of perturbation theory 
now becomes critical as it can affect the high-$n$-point amplitudes already at leading order in the weakly coupled perturbation theory.
It was pointed out in \cite{Cornwall,Goldberg} that the factorial growth could arise from the large numbers of Feynman diagrams contributing
to the scattering amplitude ${\cal A}_{n}$ at large $n$. This line of reasoning should be robust in any quantum field theory which does not 
exhibit significant cancellations between the diagrams in computations of on-shell quantities. In particular, in the scalar field theory with $\lambda \phi^4$-type
interactions tree graphs all have the same sign, and the leading-order high-multiplicity amplitudes acquire the factorial behaviour, ${\cal A}_{n} \sim \lambda^{n/2}\, n!$ which 
(assuming that the amplitudes do not decrease very rapidly in moving off the threshold) leads to
the factorial growth of the cross-section, $\sigma_n^{\rm tree} \sim \lambda^n\, n! \times f_n({\rm kinematics})$,
in violation of unitarity. This signals the breakdown of perturbative description of these observables for $n > 1/\lambda$.

\medskip

Multi-particle amplitudes in scalar field theory were studied in depth in the literature in 1990's, see
papers~\cite{Voloshin64,Argyres73,Brown,Voloshin80,Voloshin83,Smith84,LRST,GorskyV,Son,BLT,LRT} and references therein.
In section 2 we will briefly review their results for the tree-level amplitudes at threshold,
relevant for us in this paper. We will also discuss aspects of multi-boson production 
away from the threshold and the role of the loop corrections in the Conclusions section.

\medskip 

The factorial growth of the amplitudes in scalar QFT is to be contrasted with their behaviour in gauge theory. In the massless gauge theory, for example in QCD, 
the gauge invariance, on-shell conditions and other symmetries result 
in dramatic cancellations between Feynman diagrams for on-shell quantities such as scattering amplitudes. As the result there is no manifest factorial growth 
in amplitudes. For example the maximal helicity violating  (MHV) amplitudes (those with 2 negative and $n-2$ positive helicity gluons)
are given for any n by the single-term Parke-Taylor expression \cite{PT},
\[
{\cal  A}_n^{\rm tree\, MHV} \,=\, \frac{\langle r s\rangle^4}{\langle 12\rangle \langle 23\rangle \ldots \langle n1\rangle}\,,
\]
where the two negative-helicity gluons have momenta $p_r$ and $p_s$; all others are of positive helicity, and $\langle i j\rangle$ is the familiar 
holomorphic spinor product, see e.g. \cite{MP,Dixon} for a review of the spinor helicity formalism.

\medskip

The main motivation of this paper is to answer the question of whether the weak sector of the Standard Model (SM), as approximated by the spontaneously broken SU(2) gauge theory, 
will retain the factorial growth of the multi-particle tree-level amplitudes found in the scalar QFT, or will it follow the regular behaviour of QCD amplitudes.
The Gauge-Higgs theory will be analysed in section 3 using the generalisation of the Brown's generating function technique \cite{Brown} of section 2. 

\medskip

Another motivation for studying the high-multiplicity production in the electroweak sector at high energies is the close analogy and complementarity 
between these perturbative processes and the  instanton-like  processes over the sphaleron barrier\cite{Ringwald, McLVV, KR, KRTetc, Mattis, LRRT} 
which, if observable at the next generation of hadron colliders, 
would violate the Baryon plus Lepton ($B+L$) number in the SM.

\bigskip

\medskip
\section{Summing tree graphs at threshold in scalar $\phi^4$ theory}
\label{sec:2}
\medskip

Working at tree level amounts to switching off loop effects controlled by the Planck's constant $\bar{h}$, thus tree-level on-shell amplitudes or currents 
are essentially classical quantities. As such, they should be governed by classical equations of the system with an appropriate source term added to distinguish between
different multi-particle states. 
An elegant formalism for computing all tree-level multiboson amplitudes at threshold in terms of a classical generating function was introduced by Brown \cite{Brown}
for scalar field theory, which we will briefly review below.
Based on solving classical equations directly this approach readily bypasses summation over individual Feynman diagrams. 

The amplitude ${\cal A}_{1\to n}$ for the field $\phi$ to create $n$ particles in the real scalar field theory with the Lagrangian
\[
{\cal L}_\rho(\phi) \,= \, \frac{1}{2} \left(\partial \phi\right)^2 - \frac{1}{2} M^2 \phi^2 - \frac{1}{4} \lambda \phi^4
 +\, \rho\, \phi\,,
\label{eq:Lphi}
\]
is derived by differentiating the matrix element of the initial state, $\langle 0_{\rm out} |\phi(0)| 0_{\rm in}\rangle_\rho$ 
with respect to the source $\rho(x)$ and applying the LSZ reduction,
\[
\langle n|\phi(x)| 0\rangle \,=
\lim_{\rho\to0}\,
\left[\prod_{j=1}^{n}\,\lim_{p_j^2\to M^2}\int d^4 x_j e^{ip_j \cdot x_j}(M^2-p_j^2) \frac{\delta}{\delta\rho(x_j)}\right]
\langle 0_{\rm out} |\phi(x)| 0_{\rm in}\rangle_\rho\,.
\label{eq:gfunphi}
\]
Tree-level approximation is obtained by replacing the matrix element $\langle 0_{\rm out} |\phi(x)| 0_{\rm in}\rangle_\rho \longrightarrow \phi_{\rm cl}(x)$
by a solution to the classical field equation corresponding to the Lagrangian ${\cal L}_\rho(\phi)$, including the source $\rho(x)$ term.
This defines the classical field as the functional of the source, $\phi_{\rm cl}[\rho]$.

Next step is to go to the threshold limit where all the outgoing particles are produced at rest, $\vec{p}_j=0$. In this limit, it is sufficient to
consider the spatially-independent source $\rho(t)$. Specifically, before taking the $p_j^2\to M^2$ limit  in \eqref{eq:gfunphi}, we set
all outgoing momenta to $p^{\mu}_j=(\omega, \vec{0})$, and choose $\rho(t)=\rho_0(\omega)\,e^{i\omega t}$. This amounts to the substitution
\[
(M^2-p_j^2) \frac{\delta}{\delta\rho(x_j)} \longrightarrow (M^2-\omega^2) \frac{\delta}{\delta\rho(t_j)} = \frac{\delta}{\delta z(t_j)} \,,
\]
where we defined
\[
z(t)\,:=\, \frac{\rho_0(\omega)\,e^{i\omega t}}{M^2-\omega^2 -i\epsilon}\, :=\, z_0\,\,e^{i\omega t} \,.
\]
When the classical solution $\phi_{\rm cl}$ is re-expressed as the functional of the new variable $z(t)$ rather than the original source $\rho(t)$,
one can take the required on-shell limit $\omega \to M$ simultaneously with
sending the amplitude of the source to zero, $\rho_0(\omega) \to 0$ such that $z_0$ remains finite \cite{Brown}.
For each external leg
operator acting on $\phi_{\rm cl}$ in \eqref{eq:gfunphi} we have,
\[
\int d^4 x_j e^{ip_j \cdot x_j}(M^2-p_j^2) \frac{\delta}{\delta\rho(x_j)} \, \cdot\, \phi_{\rm cl}(t)
\,=\, e^{i\omega t}\, \frac{\partial \phi_{\rm cl}}{\partial z(t)}
\,=\,  \frac{\partial \phi_{\rm cl}}{\partial z_0}\,.
\label{eq:2gfunphi}
\]
The tree-level amplitude ${\cal A}_{1\to n}$ at the $n$-particle threshold is thus given by
\[
{\cal A}_{1\to n}\,=\, \langle n|\phi(0)| 0\rangle \,=\,
\left.\left(\frac{\partial}{\partial z}\right)^n \phi_{\rm cl} \,\right|_{z=0}
\,,
\label{eq:ampln1}
\]
where the generating function $\phi_{\rm cl} (z(t))$ is a particular classical solution which we will now determine.
As we already noted, $\phi_{\rm cl} (z(t))$ is unaffected by the double scaling limit, $\omega \to m$ with $\rho_0(\omega) \to 0$,
and at the same time the source term drops out from its defining classical equation.
It reduces to an ordinary differential equation for $\phi(t)$ with no source term. For the theory described by the Lagrangian \eqref{eq:Lphi}
it reads,
\[
d_t^2\phi + M^2 \phi +\lambda\phi^3 \,=\, 0\,.
\label{cleqp4}
\]
To give the generating function of amplitudes at multiparticle thresholds, the solution must contain only the positive frequency 
components of the form $e^{+ i n M t}$ where $n$ is the number of final state particles in the amplitude ${\cal A}_{1\to n}$. This follows immediately from \eqref{eq:ampln1}.
Thus, the solution we are after is given by the Taylor expansion in powers of the complex variable $z(t)$,
\[
\phi_{\rm cl} (t) \,=\, z(t) \,+\, \sum_{n=2}^{\infty} d_n\, z(t)^n
\label{gen-fun1}
\]
In the limit where interactions are switched off, $\lambda=0$, the correctly normalised solution is $\phi_{\rm cl} = z(t)$ and this fixes the coefficient of the
first term on the {\it r.h.s.} of \eqref{gen-fun1}.  
As the solution contains only positive frequency harmonics, it is a complex function of Minkowski time. This also fixes the initial conditions of the 
solution, $\phi_{\rm cl} (t) \to 0$ as Im$(t)\to \infty$. In Euclidean time the solution is real. 

Coefficients $d_n$ determine the actual amplitudes via \eqref{eq:ampln1},
\[
{\cal A}_{1\to n}\,=\, n! \, d_n\,,
\label{Amps-s}
\]
they can either be read off the classical soluton when it is known, or otherwise be found directly by solving equations of motion iteratively in powers of $z$.

\medskip
The classical generating function approach of \cite{Brown} amounts to finding the $\vec{x}$-independent solution of the Euler-Lagrange equations
as an analytic function of $z$ in the form \eqref{gen-fun1}, and computing the amplitudes via \eqref{eq:ampln1} or \eqref{Amps-s}.
\medskip

The classical generating function for the theory defined by \eqref{eq:Lphi} is surprisingly simple and can be written in closed form \cite{Brown},
\[
\phi_{\rm cl} (t) \,=\, \frac{z(t)}{1-\frac{\lambda}{8M^2}\, z(t)^2}\,.
\label{noSSB}
\]
It is easily checked that the expression in \eqref{noSSB} solves the classical equation \eqref{cleqp4} and has the correct form, $\phi_{\rm cl} = z + \ldots$ as $z\to 0$.
Scattering amplitudes at threshold are then given by \cite{Voloshin64,Brown}
\[
{\cal A}_{1\to n}\,=\, 
\left.\left(\frac{\partial}{\partial z}\right)^n \phi_{\rm cl} \,\right|_{z=0}
\,=\, n!\, \left(\frac{\lambda}{8M^2}\right)^{\frac{n-1}{2}}
\,.
\label{eq:ampln2}
\]
We see that the leading order (tree-level) amplitudes in scalar QFT grow factorially with $n$, which ultimately is the consequence of the growth in the number of diagrams in perturbation theory.

\medskip

The generating functions formalism works equally well also in the 
 real scalar field theory with spontaneously broken ${\cal Z}_2$ symmetry. 
 For future reference we will denote the scalar field of this model as $h(x)$.
 The Lagrangian takes the familiar form,
 \[
{\cal L}(h) \,= \, \frac{1}{2} \left(\partial h\right)^2 - \frac{\lambda}{4} \left( h^2 - v^2\right)^2
\,,
\label{eq:LSSB}
\]
where $v$ is the VEV of $h(x)$.
The classical equation for the spatially uniform field $h(t)$,
\[
d_t^2 h \,=\, -\lambda\,h^3 +\lambda v^2\,h
\,,
\label{cleq-SSB}
\]
again has a simple closed-form solution \cite{Brown}:
\[
h_{\rm cl} (t) \,=\, v\,\frac{1+\frac{z(t)}{2v}}{1-\frac{z(t)}{2v}} \, , \quad {\rm where} \quad
z(t)\,=\,z_0 \, e^{iM_h t}\, = \, z_0 \, e^{i\sqrt{2\lambda}\,v\,t}
\label{sol-SSB}
\]
with the correct form of initial conditions,  $h_{\rm cl} = v + z +\ldots $ as $z\to 0$.
The Taylor expansion of the generating function \eqref{sol-SSB} can be recast in the form ({\it cf.} \eqref{gen-fun1}),
\[
h_{\rm cl} (t) \,=\, 2v\, \sum_{n=0}^{\infty} \left(\frac{z(t)}{2v}\right)^n d_n 
\,=\, v\,+\, 2\sum_{n=1}^{\infty} \left(\frac{z(t)}{2v}\right)^n
\, ,
\label{gen-funh}
\]
i.e. with $d_0=1/2$ and all $d_{n\ge1}=1$. Scattering amplitudes at threshold in this theory are given by
\cite{Argyres73,Brown}
\[
{\cal A}_{1\to n}\,=\, 
\left.\left(\frac{\partial}{\partial z}\right)^n h_{\rm cl} \,\right|_{z=0}
\,=\, n!\, (2v)^{1-n}
\,.
\label{eq:amplnh}
\]
This theory of a single real scalar field with a spontaneously broken $h\to-h$ discrete symmetry is a toy version of the
Higgs sector of the SM. The field $h$ looks lie the SM Higgs in the unitary gauge, but without the accompaniment of longitudinal massive vector bosons.
In the following section we will apply the generating functions method to the spontaneously broken gauge theory and examine if 
the factorial growth, which is manifest in \eqref{eq:amplnh}, also persists in the Gauge-Higgs theory.

\medskip
\section{Multiparticle production in the Gauge-Higgs theory}
\label{sec:3}
\medskip

For concreteness we will concentrate on the
simplest Non-Abelian case of interest --  the SU(2) gauge theory spontaneously broken by the vacuum expectation value $v$ of the Higgs doublet,
\[
{\cal L}\,=\, -\frac{1}{4} F^{a\,\mu\nu}F^a_{\mu\nu} \,+\, |D_\mu H|^2 \,-\, 
\lambda\left(|H|^2-\frac{v^2}{2}\right)^2\,.
\label{eq:LGH}
\]
This theory describes the weak sector of the SM in the limit of the vanishing $\theta_{\rm W}$.
In the unitary gauge,
\[
H= \frac{1}{\sqrt{2}} \left(0,h\right),\,
\]
and the Higgs potential in terms of $h$ takes the same form as in Eq.~\eqref{eq:LSSB}. The particle content of the model is given by the neutral Higgs state,
$h$, and a triplet of massive vector bosons, $W^{\pm}$ and $Z^0$,
described by $A_\mu^a$ with $a=1,2,3$, which we will collectively refer to as $V$. The Higgs mass and the mass of the vector boson triplet are given by,
\[
M_h\,=\, \sqrt{2\lambda}\,v \,\simeq\, 125.66\, {\rm GeV}\,, \qquad
M_V\,=\, \frac{g v}{2}\,\simeq\, 80.384 \, {\rm GeV}\,,
\label{eq:masses}
\]
where we have also shown their numerical values, set by the SM Higgs and $W$ boson masses, which will be uses in our calculations of the amplitudes below.

The kinematic regime of interest is when a single virtual state -- the Higgs or a gauge boson, decays into 
$n$ Higgs bosons and $m$ vector bosons, all with vanishing spatial momenta. In the rest frame of the initial virtual boson we have
\[
p_{\rm in}^\mu=\,(nM_h + mM_V, \vec{0}) \,\rightarrow \,\sum_{j=1}^{n} p_j^\mu\,+\,\sum_{k=1}^{m} p_k^\mu
\,,\quad {\rm where}\quad p_j^\mu=(M_h, \vec{0})\,,\quad p_k^\mu=(M_V, \vec{0})\,.
\label{eqn:thrp}
\]
This system can be Lorentz boosted, giving all momenta a non-vanishing $p^3$ component. The process in this frame corresponds
to the highly virtual boson being produced in the $pp$ collision (e.g. the gluon gluon fusion to a Higgs) which decays to a maximal kinematically allowed number 
of Higgses and massive vector bosons.
The boosted direction along  $p^3$  is the longitudinal direction. All momenta in the boosted frame have the form, $p^\mu = (p^0,0,0,p^3)$. 

For the rest of the analysis we return to the rest frame \eqref{eqn:thrp} where $p^3$ is vanishing (or infinitesimally small). 
The transversality condition, $p^\mu A_\mu^a=0$, allows us to set $A_0^a=0$. We thus are left with the following degrees of freedom:
$\{h(t),A_m^a(t)\}$ with $m=1,2,3$, the first two being the transverse and the third -- the longitudinal polarisations of the triplet ($a=1,2,3$) of massive vector bosons.

The Lagrangian \eqref{eq:LGH} reduced on these spatially-independent components reads, 
\[
{\cal L}\,=\, \frac{1}{2} (d_t A_m^a)^2 + \frac{1}{2} (d_t h)^2 -  \frac{g^2}{8} h^2 (A_m^a)^2 
- \frac{g^2}{4}  \left((A_m^a)^2 (A_n^b)^2 -(A_m^a A_n^a)^2\right)
- \frac{\lambda}{4}\left(h^2-v^2\right)^2
\label{eq:LGH2}
\]
and the equations of motion for $h_{\rm cl}(t)$ and $A_{m\,\rm cl}^a(t)$ are,
\begin{eqnarray}
d_t^2 h &=& -\lambda\,h^3 +\lambda v^2\,h-\frac{g^2}{4} (A_m^a)^2 h
\,,\label{cleq-h}\\
d_t^2 A_m^a &=& -\frac{g^2}{4} h^2 A_m^a - g^2 \left((A_n^b)^2 A_m^a -(A_n^a A_n^b) A_m^b\right)
\label{cleq-A}
\end{eqnarray}
This system of equations can now be solved iteratively. 

To simplify the derivation we will assume that the final state does not contain transverse polarisations of the vector bosons,
and concentrate on the production of longitudinal polarisations, $A_3^a$ and Higgses $h$. The `commutator' term in 
\eqref{eq:LGH2} and \eqref{cleq-A}
then drops out and we get,
\begin{eqnarray}
d_t^2 h &=&   -\lambda\,h^3  +\lambda v^2\,h -\frac{g^2}{4} (A_L^a)^2 h
\,,\label{cleq-h3}\\
d_t^2 A_L^a &=& -\frac{g^2}{4} h^2 A_L^a \,.
\label{cleq-A3}
\end{eqnarray}
The classical solutions required to give the generating function of the amplitudes on the multi-$h$, multi-$V_L$ threshold,
should be the analytic functions (i.e. given by the double Taylor expansion in terms) of two variables,
\[
z(t) \,=\, z_0\, e^{i M_h t}\,,\qquad 
w^a(t)\,=\, w_0^a\, e^{i M_V t}\,,
\label{eq:zw}
\]
with the leading-order terms being,
\[
h_{\rm cl}(t)\,=\, v + z(t) + \ldots\,,\qquad 
A_{L\,\rm cl}^a (t)\,=\, w^a(t) + \ldots\,,
\label{eq:leadzw}
\]
The system of equations \eqref{cleq-h3}-\eqref{cleq-A3} can be shown to depend only on a single parameter,
by re-writing them in terms of dimensionless variabeles, $t=M_h t$, $h=h/v$ and $A_L^a= A_L^a/v$,
\begin{eqnarray}
d_t^2 \,h &=& -\frac{1}{2}(h^3 -h) - \kappa^2 (A_L^a)^2 h
\,,\label{cleq-h3r}\\
d_t^2 A_L^a &=& -\kappa^2 h^2 A_L^a \,,
\label{cleq-A3r}
\end{eqnarray}
with 
\[
\kappa \,:=\, \frac{g}{2\sqrt{2\lambda}} \,=\,  \frac{M_V}{M_h}\,.
\label{kappadef}
\]
Before we write down the double Taylor expansion of the generating functions $h_{\rm cl}$ and $A_{L\,\rm cl}^a$, we define the
scalar function ${\rm A}_{\rm cl}$ for vector bosons via
\[
A_{L\,\rm cl}^a = w^a\, {\rm A}_{\rm cl}\,,
\]
and introduce the new combination,
\[ W=w^aw^a\,.
\]
We now write in the double Taylor expansion in the form,
\begin{eqnarray}
h_{\rm cl} (z,W) &=& \sum_{n=0}^{\infty}\sum_{k=0}^{\infty} \, d_{n,k}\,z^n\,  W^k\,,\quad {\rm with}\,\, d_{0,0}=1\,{\rm and}\, d_{1,0}=1\,,
\label{eq:dTh}
\\
{\rm A}_{\rm cl}(z,W) &=& \sum_{n=0}^{\infty}\sum_{k=0}^{\infty} \, a_{n,k}\,z^n\,  W^k\,,\quad {\rm with}\,\, a_{0,0}=1\,.
\label{eq:dTA}
\end{eqnarray}
Differentiating these expressions twice with $t$ we write down the
equations \eqref{cleq-h3r}-\eqref{cleq-A3r} in the form\footnote{Recall that in terms of our dimensionless variables, 
$z(t)=z_0 \, e^{i t}$, $w(t)=w_0 \, e^{i \kappa t}$ and $W=w_0^2 \, e^{i 2 \kappa t}$.},
\begin{eqnarray}
d_{n,k} (n+2\kappa \,k)^2 \,z^n\,  W^k &=& \left.\left[\frac{1}{2}(h^3 -h) + \kappa^2 \,W\,{\rm A}_{\rm cl}^2 h\right]\right|_{z^n W^k}
\,,\label{cleq-h3rT}\\
a_{n,k} (n+ \kappa+ 2\kappa \,k)^2 \,z^n\,  W^k &=& \left.\left[\kappa^2 h^2 \,{\rm A}_{\rm cl} \right]\right|_{z^n W^k}\,.
\label{cleq-A3rT}
\end{eqnarray}

\medskip

These equations are solved iteratively as follows. First we set $k=0$ and solve the Higgs equations \eqref{cleq-h3rT}
for all values of $n$ thus determining all coefficients\footnote{Not surprisingly, we find
$d_{n\ge1,0}= 1/2^{n-1}$ in accordance with \eqref{gen-funh}.}
 $d_{n,0}$. No other coefficients enter this equation for $k=0$. Then we solve the $A$-equations \eqref{cleq-A3rT}
 for the coefficients $a_{n,0}$ for each $n$.
 Next we set $k=1$, and solve equations \eqref{cleq-h3rT} for all $n$ to determine $d_{n,1}$. Following this, the coefficients
 $a_{n,1}$ are found by solving  \eqref{cleq-A3rT} at $k=1$ for all values of $n$. This procedure is repeated for all values of $k$.
  
 After implementing this iterative algorithm in {\it Mathematica} we can solve for $d_{n,k}$ and $a_{n,k}$ to any desired values of
 $n$ and $k$ numerically.
To make the results described in this paper more useful and readily available to researchers, we have prepared a Mathematica notebook  included with the submission files for this paper on the arXiv. It can also be obtained here http://tinyurl.com/lj6m53u
 Ref.~\cite{Math}.
 
 \medskip

There is not much hope to find a simple analytical solution for the generating functions as was the case in the scalar field theory;
even at $k=0$ the $a_{n,0}$ coefficients start becoming increasingly complicated already at relatively low values of $n$, 
\begin{eqnarray}
a_{4,0} &=& \frac{\kappa^2 (9 + 33 \kappa + 75 \kappa^2 + 90 \kappa^3 + 64 \kappa^4 + 24 \kappa^5 + 4 \kappa^6)}{24 (1 + 
   \kappa) (2 + \kappa) (1 + 2 \kappa) (3 + 2 \kappa)} \, \nonumber\\
a_{5,0} &=& 
\frac{\kappa^2 (90 + 375 \kappa + 987 \kappa^2 + 1500 \kappa^3 + 1474 \kappa^4 + 920 \kappa^5 + 
   360 \kappa^6 + 80 \kappa^7 + 8 \kappa^8)}{120 (1 + \kappa) (2 + \kappa) (1 + 2 \kappa) (3 + 
   2 \kappa) (5 + 2 \kappa)} \,\nonumber
 \end{eqnarray}
 However,  the closed form solution is not really needed in order to determine amplitudes at threshold as they are described by the coefficients $d_{n,k}$ and $a_{n,k}$
 which are computed straightforwardly in our iterative procedure, as described above.
 
 Before we list the coefficients we solved for, we re-write the double Taylor expansion of the generating functions back in terms
 of physical dimensionful variables, with an additional rescaling by factors of 2 -- to make the comparison with Eq.~\eqref{gen-funh}
 more manifest. We define the rescaled coeffiecients
 \[ d(n,2k) \,:=\, 2^{n+2k-1}\, d_{n,k}\,,\qquad
 a(n,2k) \,:=\, 2^{n+2k}\, a_{n,k}\,,
 \]
 and write down the double Taylor expansion of the generating functions in terms of these as follows,
 \begin{eqnarray}
h_{\rm cl} (z,w^a) &=& 2v\, \sum_{n=0}^{\infty}\sum_{k=0}^{\infty} \, d(n,2k)\,\left(\frac{z}{2v}\right)^n\,\left(\frac{w^a w^a}{(2v)^2}\right)^k
\,,\label{eq:dTh-fin}
\\
A_{L\,\rm cl}^a(z,w^a) &=& w^a\,\sum_{n=0}^{\infty}\sum_{k=0}^{\infty} \, a(n,2k)\, \left(\frac{z}{2v}\right)^n\,\left(\frac{w^a w^a}{(2v)^2}\right)^k
\,,
\label{eq:dTA-fin}
\end{eqnarray}
where $z(t)$ and $w^a(t)$ are given by Eqs.~\eqref{eq:zw}.   

\begin{figure}[]
\begin{center}
\vspace*{-1.cm}
\includegraphics[width=1.\textwidth]{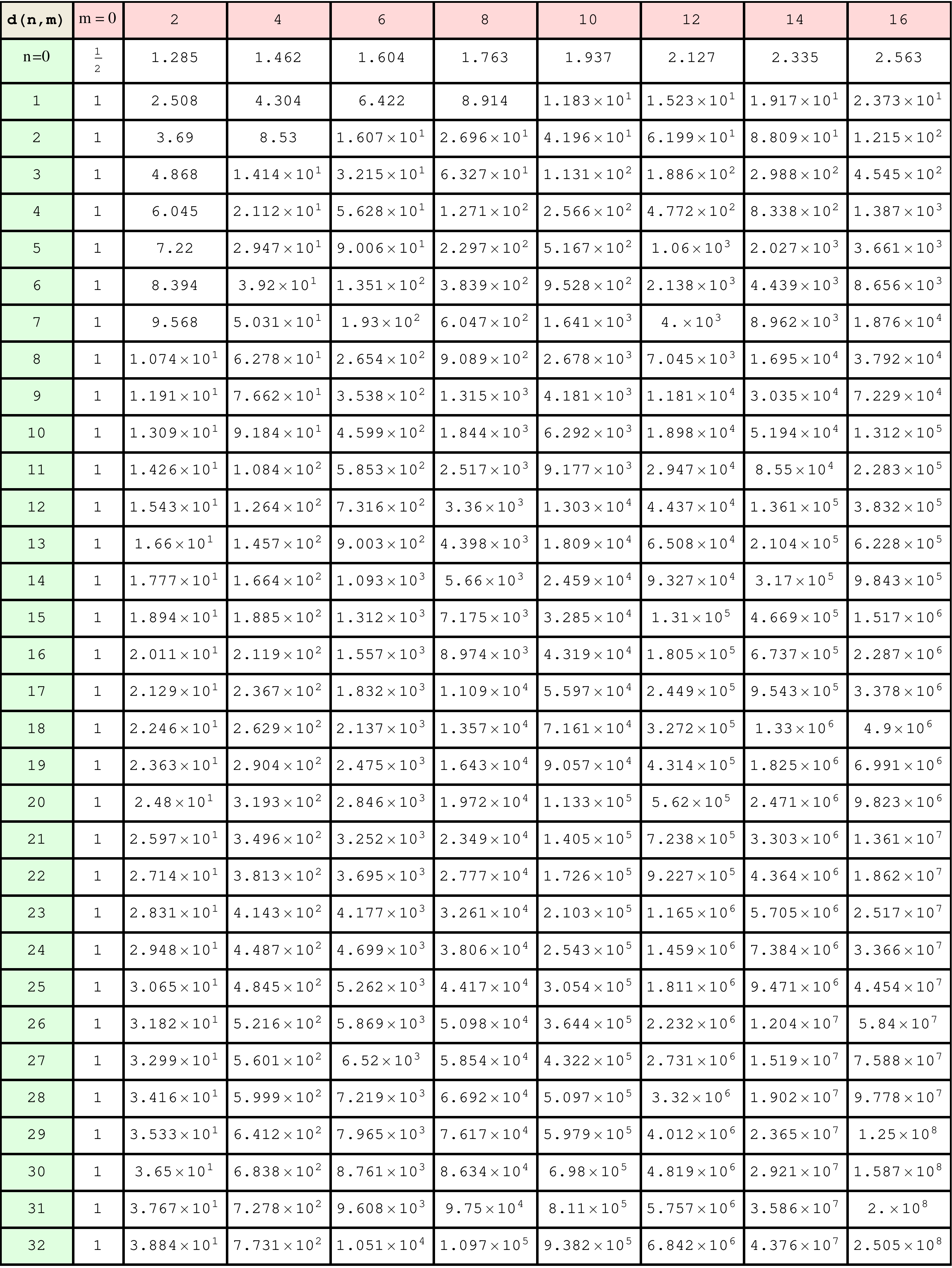}
\caption{Coefficients $d(n,m)$ for $n=0\ldots 32$ and $m=0\ldots 16$ of the virtual Higgs decay amplitudes 
${\cal A}(h \to n\times h + m\times V_L)= \, n!\, m!\, d(n,m) \, (2v)^{1-n-m}$ 
at the $(n+m)$-particle threshold. $V_L$ are the longitudinal components of the $W^{\pm},Z^0$ massive SU(2) vector bosons. 
The value of $\kappa$ is set to the SM value, $\kappa= M_W/M_h= 80.384/125.66\simeq 0.6397$. 
Mathematica implementation is in \cite{Math}.}
\label{Tab:1}
\end{center}
\end{figure}

\begin{figure}[]
\begin{center}
\vspace*{-1.cm}
\includegraphics[width=1.\textwidth]{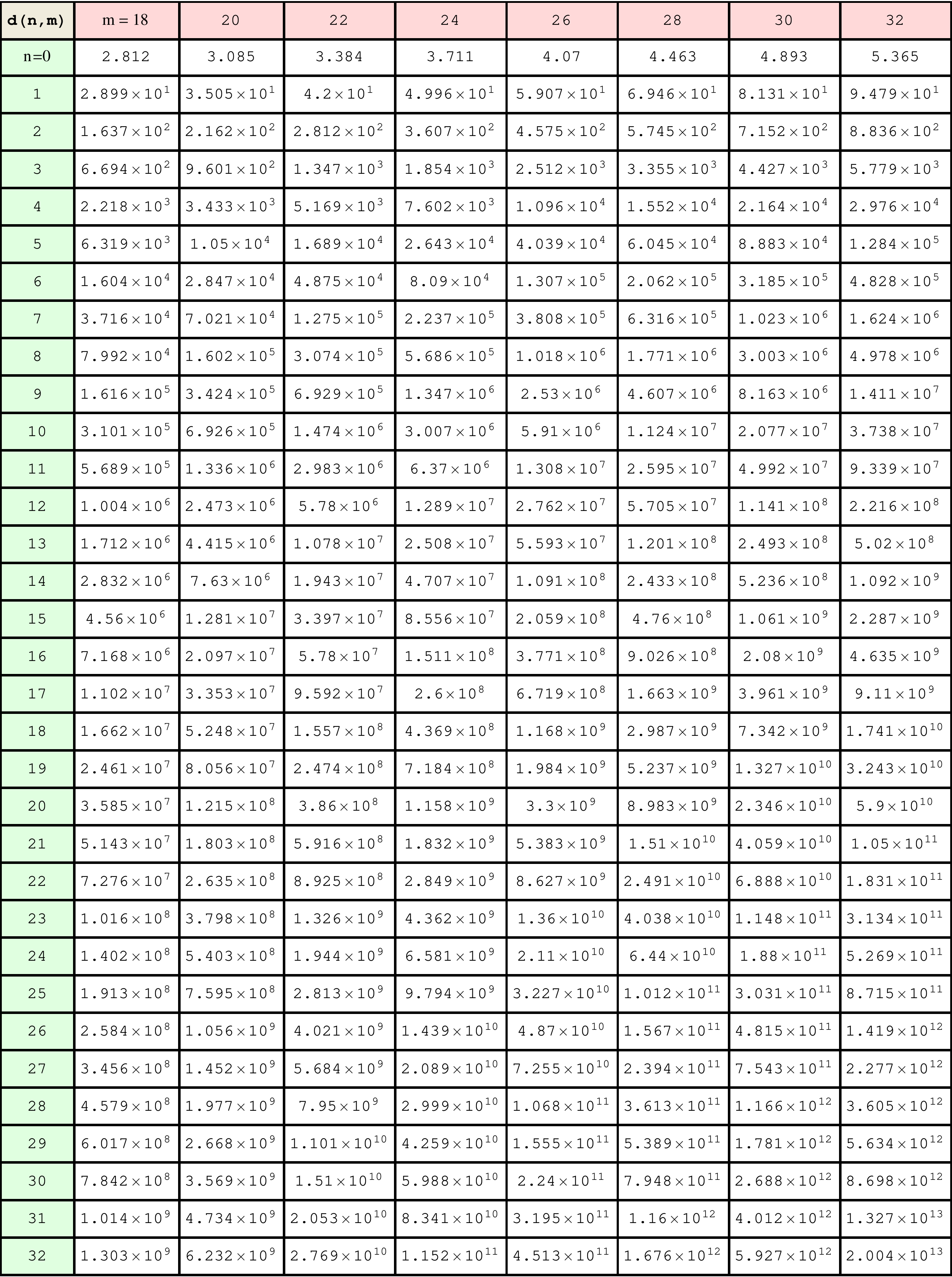}
\caption{Second half of the coefficients $d(n,m)$ for $m=16\ldots 32$ of the virtual Higgs decay amplitudes 
${\cal A}(h \to n\times h + m\times V_L)= \, n!\, m!\, d(n,m) \, (2v)^{1-n-m}$ 
at the $(n+m)$-particle threshold. $\kappa= M_W/M_h= 80.384/125.66\simeq 0.6397$.
Mathematica implementation is in \cite{Math}.}
\label{Tab:2}
\end{center}
\end{figure}

\begin{figure}[]
\begin{center}
\vspace*{-1.cm}
\includegraphics[width=1.\textwidth]{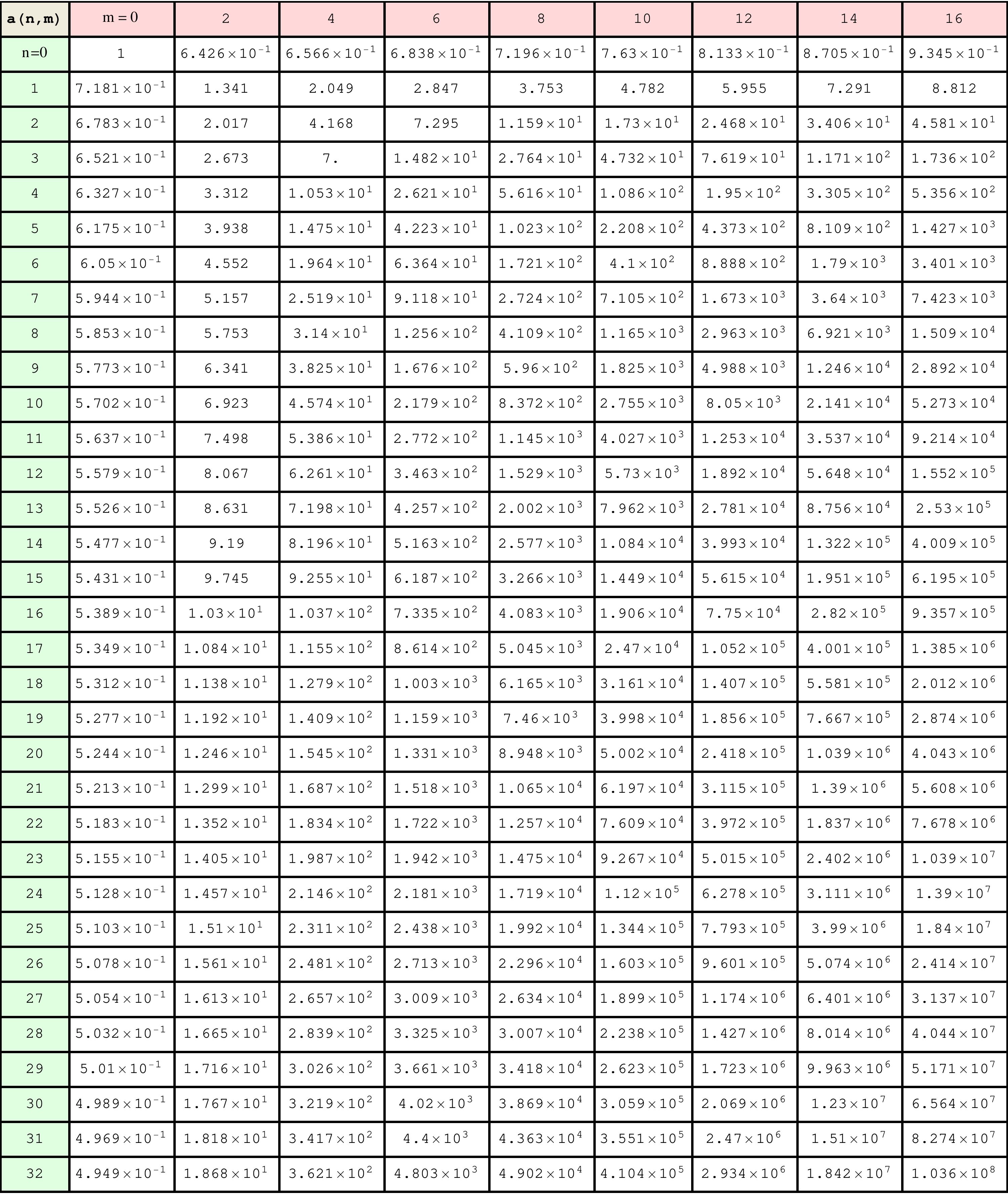}
\caption{Coefficients $a(n,m)$ for $n=0\ldots 32$ and $m=0\ldots 16$ of the virtual vector boson decay amplitudes 
${\cal A}(V_L \to n\times h + (m+1)\times V_L)= \, n!\, (m+1)!\, a(n,m) /(2v)^{n+m}$ 
at the $(n+m+1)$-particle threshold. 
The value of $\kappa$ is set to the SM value, $\kappa= M_W/M_h= 80.384/125.66\simeq 0.6397$.
Mathematica implementation is in \cite{Math}.}
\label{Tab:3}
\end{center}
\end{figure}

\begin{figure}[]
\begin{center}
\vspace*{-1.cm}
\includegraphics[width=1.\textwidth]{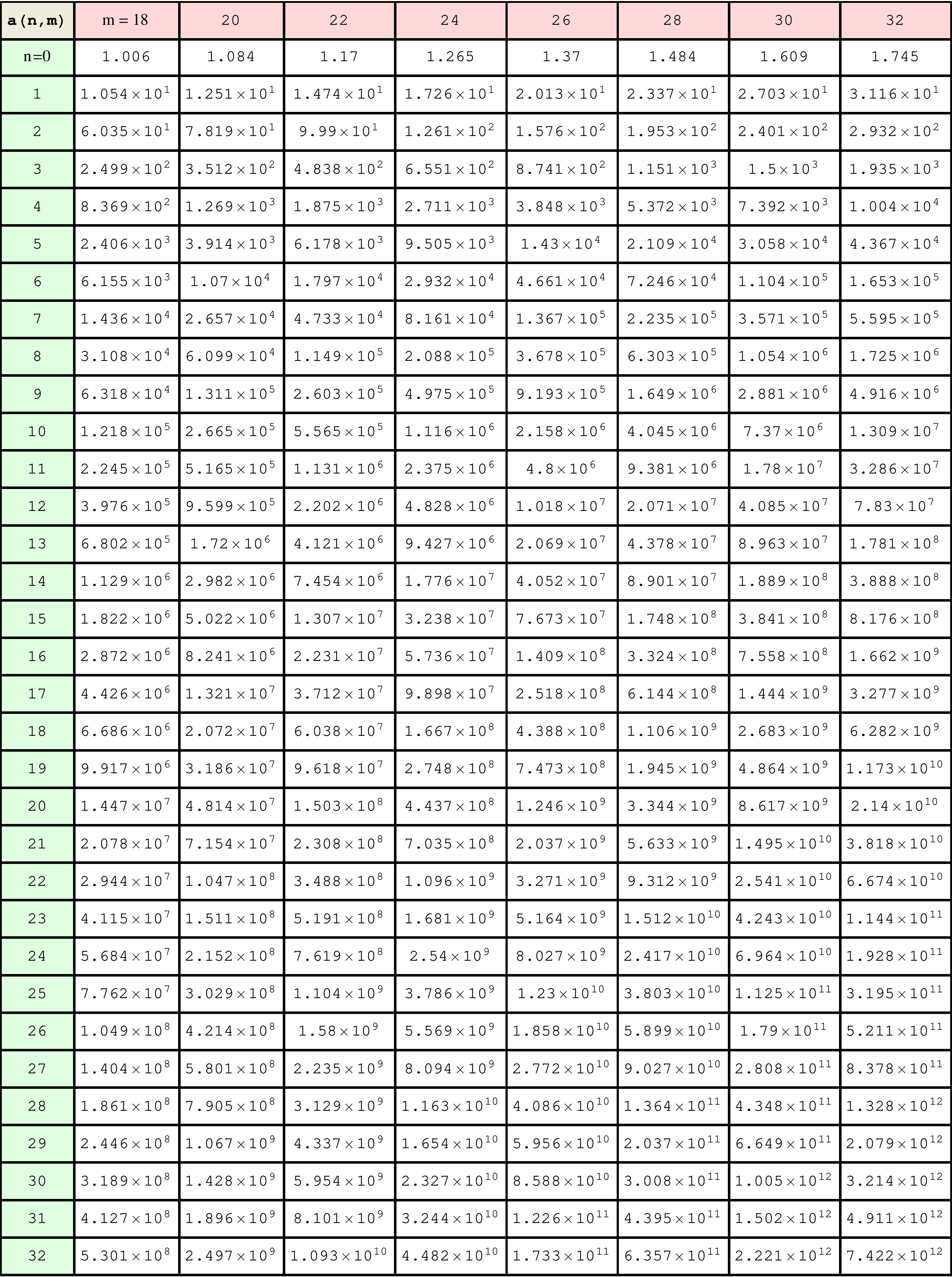}
\caption{Second half of the coefficients $a(n,m)$ for $m=16\ldots 32$ 
of the virtual vector boson decay amplitudes 
${\cal A}(V_L \to n\times h + (m+1)\times V_L)$ 
at the $(n+m+1)$-particle threshold. As before, $\kappa= M_W/M_h= 80.384/125.66\simeq 0.6397$.
Mathematica implementation is in \cite{Math}.}
\label{Tab:4}
\end{center}
\end{figure}

The scattering amplitudes on multiparticle thresholds are obtained by repeatedly differentiating the generating functions in \eqref{eq:dTh-fin}-\eqref{eq:dTA-fin}
with respect to $z$ and $w^a$. For example, for the Higgs to $n$ Higgses and $m$ longitudinal Z bosons threshold amplitude we get,
\[
  {\cal A}(h \to n\times h + m\times Z_L)= \, (2v)^{1-n-m}\, n!\, m!\, d(n,m) \,,
  \label{Ah-fin}
 \] 
 and for the longitudinal Z decaying into $n$ Higgses and $m+1$ vector bosons we have,
 \[
 {\cal A}(Z_L \to n\times h + (m+1)\times Z_L)= \, \frac{1}{(2v)^{n+m}}\, n!\, (m+1)!\, a(n,m)\,.
  \label{AZ-fin}
 \]
 The amplitudes with all varieties of $W^{\pm}_L$ and $Z_L$ in the final state, one should simply differentiate 
 with respect to $w^a$ with the appropriate values of the isospin index $a=1,2,3$.
 
 \medskip
 
In the Tables in Figs.~\ref{Tab:1}-\ref{Tab:4} we list the values of the coefficients $d(n,m)$ and $a(n,m)$ describing the
amplitudes with up to  
$n=32$ Higgs bosons plus $m=32$ longitudinal vector bosons in the final state. In solving for these coefficients we have set
 $\kappa= M_W/M_h= 80.384/125.66\simeq 0.6397$.
 
 In our normalisation conventions where the generating functions and amplitudes include inverse powers of $2v$, the coefficients 
 of pure multi-Higgs production,
 $d(n,0)$ for all $n\ge1$, are all equal to 1, as can be seen in the first Table in Fig.~\ref{Tab:1}. This provides a useful reference point for the
 size of other generating function's coefficients.
 Note that by allowing the production of
 longitudinal vector bosons, i.e. after switching on $m >0$, the coefficients of the generating functions grow steadily with $m$, reaching $d(n,m) \sim 10^8$
 at $m\ge 16$ and $n \ge 32$; and $d(n,m) \sim 10^{13}$
 at $m=32 $ and $n =31$, {\it cf.} the Table in Fig.~\ref{Tab:1}. 
 Similar growth with $m$ occurs for the $a(n,m)$ coefficients of the gauge field generating function, see Tables in Figs.~\ref{Tab:3}-\ref{Tab:4}. This numerical growth 
 of the coefficients is
 in addition to the multiplicative $n!$ and $m!$ factors in the amplitudes in \eqref{Ah-fin}-\eqref{AZ-fin}. 
 
 \begin{figure}[]
\begin{center}
\begin{tabular}{cc}
\hspace{-.4cm}
\includegraphics[width=0.5\textwidth]{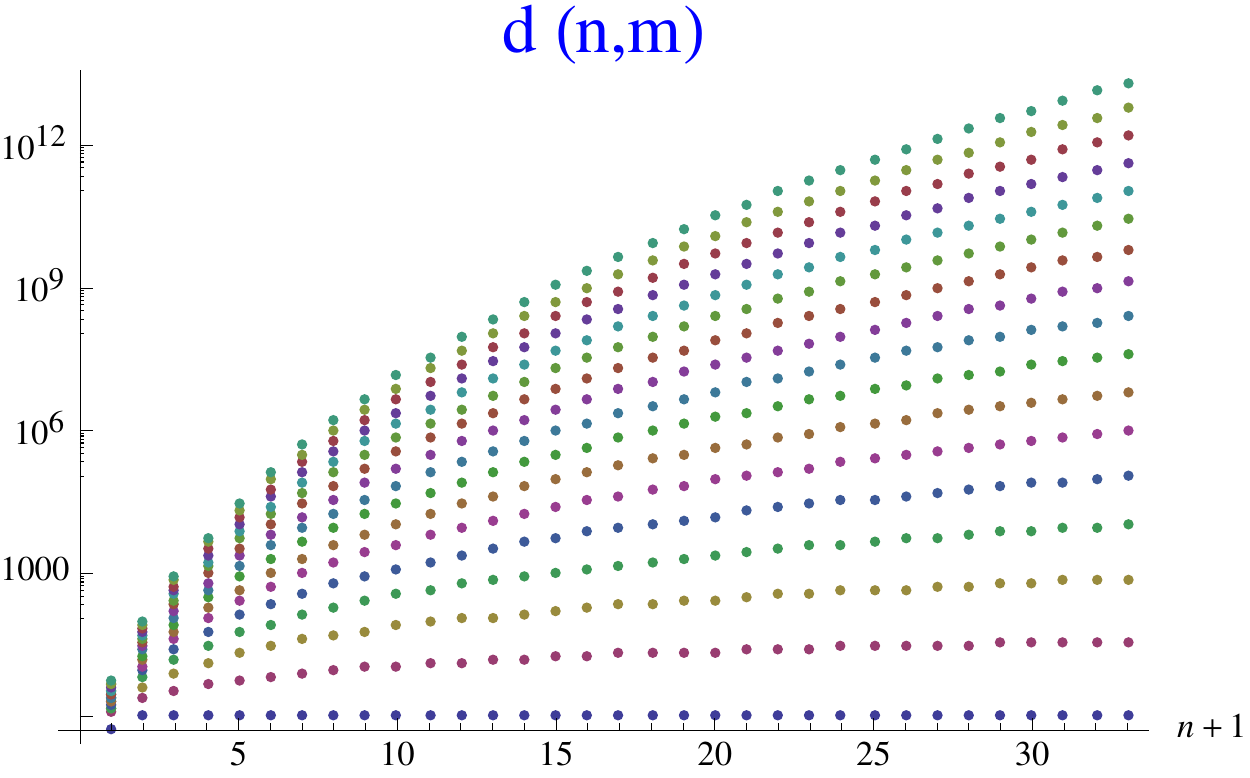}
&
\includegraphics[width=0.5\textwidth]{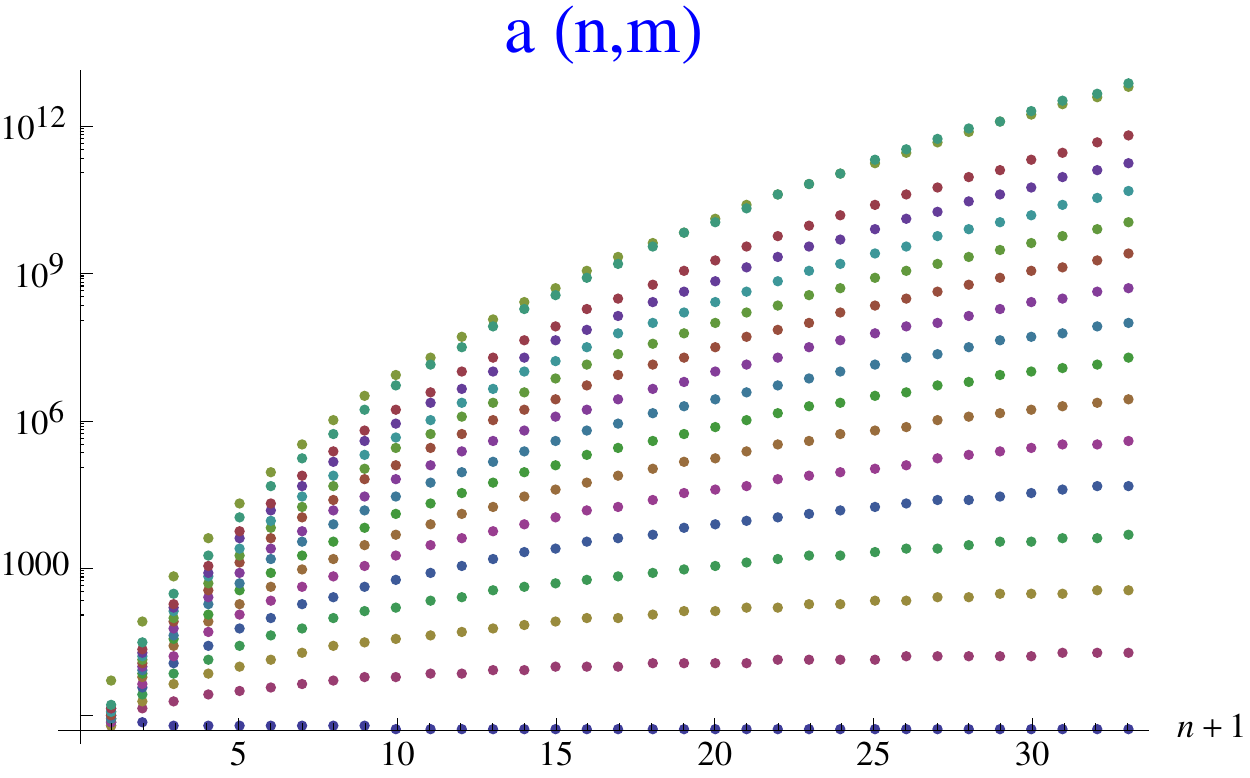}
\\
\end{tabular}
\end{center}
\vskip-.4cm
\caption{
Coefficients $d(n,m)$ and $a(n,m)$ for generating functions of amplitudes at threshold. $\kappa= M_W/M_h= 80.384/125.66\simeq 0.6397$.
 The label $n=0\ldots 32$ is shown along the horizontal axis and the sequences of curves correspond to $m=0, 2, \ldots,32$ from bottom to top.
 Mathematica implementation is in \cite{Math}.}
\label{fig:da}
\end{figure}

 In Figure~\ref{fig:da} we show the logarithmic 
 plots of all $d(n,m)$ and $a(n,m)$ for $n=0\ldots 32$ and $m=0, 2, \ldots,32$. These plots can be interpreted as sequences of curves, each curve
 representing $d(n,m)$ and $a(n,m)$ as functions of $n$ for a fixed value of $m$. Increasing values of $m=0, 2, \ldots,32$ corresponds to moving 
 upwards from lower to higher curves. 
 \begin{figure}[b!]
\begin{center}
\begin{tabular}{cc}
\hspace{-.4cm}
\includegraphics[width=0.5\textwidth]{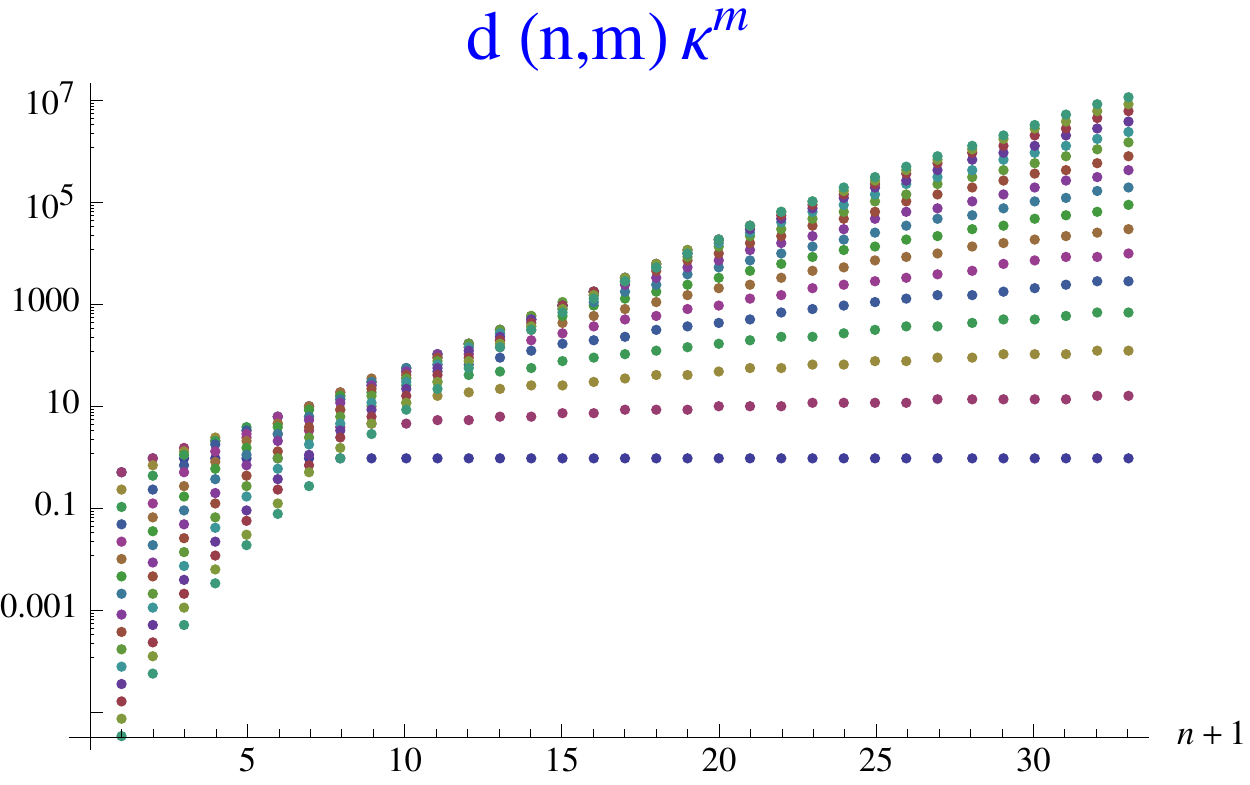}
&
\includegraphics[width=0.5\textwidth]{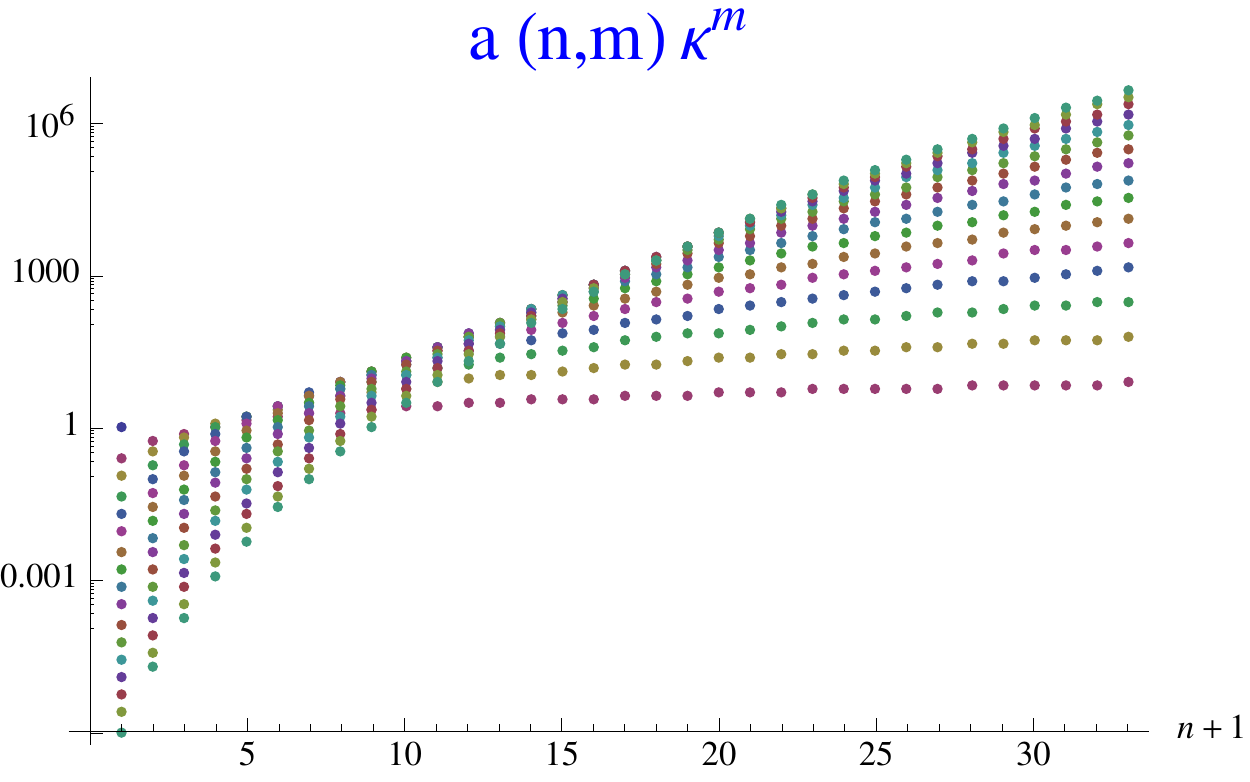}
\\
\end{tabular}
\end{center}
\vskip-.4cm
\caption{
Amplitude coefficients of Figure~\ref{fig:da}  rescaled by $\kappa^{m}$ (where $\kappa\simeq 0.6397$).}
\label{fig:daK}
\end{figure}

 For the relative comparison between the coefficients it might be more appropriate to normalise vector boson and Higgs legs by their respective masses
 rather than a universal factor of $2v$. This would be characterised by rescaling the coefficients by $\kappa^m$. These are shown in Fig.~\ref{fig:daK},
 from which we see that there is still more than six orders of magnitude growth in coefficients as one increases $m$ to $\sim 30$. 
  
\begin{figure}[]
\begin{center}
\begin{tabular}{cc}
\hspace{-.4cm}
\includegraphics[width=0.5\textwidth]{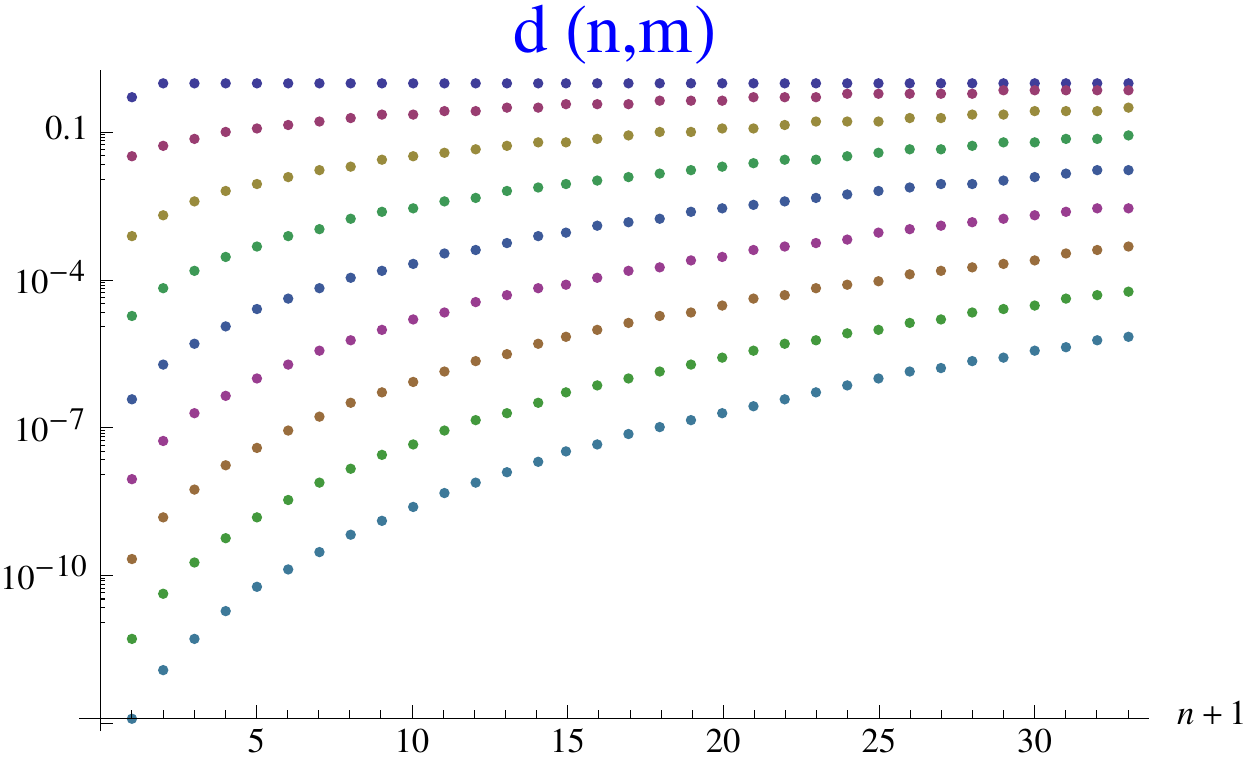}
&
\includegraphics[width=0.5\textwidth]{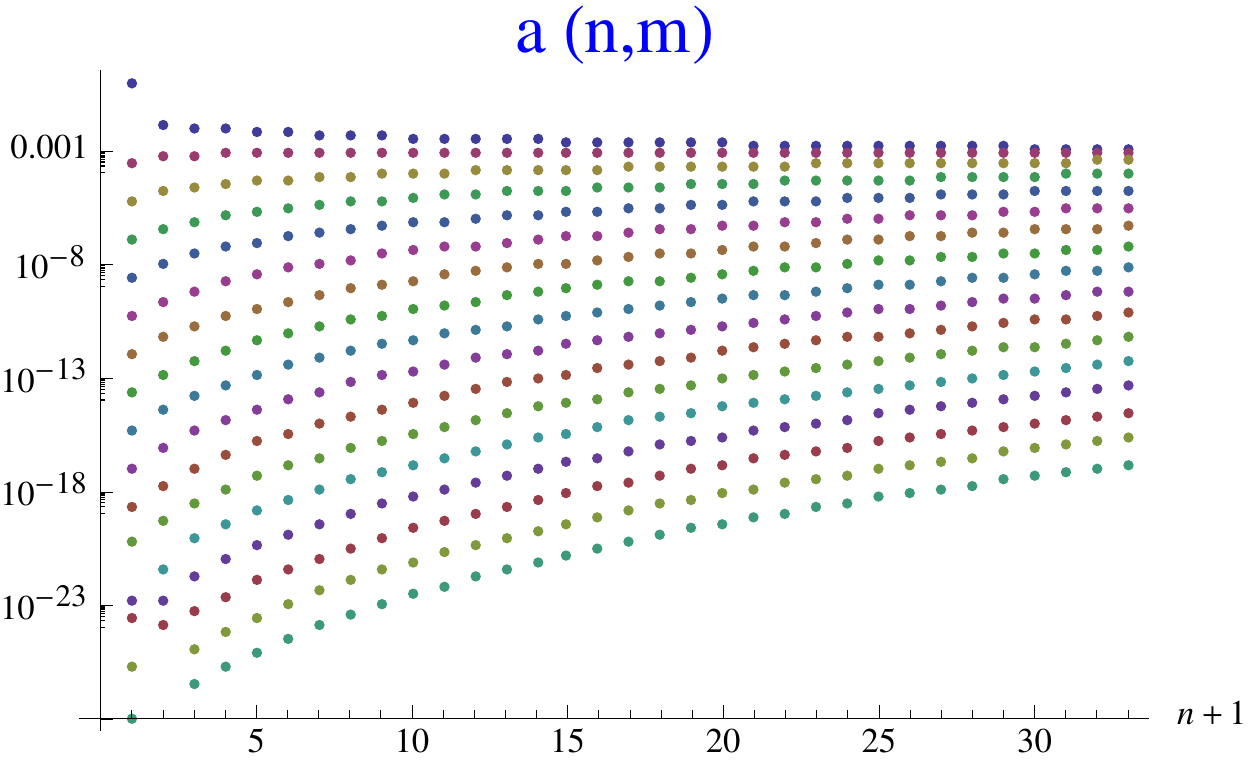}
\\
\end{tabular}
\end{center}
\vskip-.4cm
\caption{
The light vector boson case, $M_V \ll M_h$, with the non-SM choice  
$\kappa= M_V/M_h= 0.1$. The amplitude coefficients $d(n,m)$ and $a(n,m)$ are shown as functions of 
$n=0\ldots 32$. The sequences of curves correspond to $m=0, 2, \ldots,32$ from top to bottom.}
\label{fig:da10th}
\end{figure}

The importance of multiple vector boson emissions relative to the multi-Higgs production should decrease as one
decreases the vector boson mass relative to the Higgs mass, so that in the asymptotic case of $M_V=0$ we are left only
with $d(n,0)$, as the classical equations would dictate. This is indeed the case as can be seen from Fig.~\ref{fig:da10th}
which shows the (unrescaled) coefficients $d(n,m)$ and $a(n,m)$ for the case of the vector boson taken to be 10 times lighter than the Higgs.
In this case, the coefficients actually steadily decrease when one increases $m$.

\medskip
\section{Conclusions}
\label{sec:concl}
\medskip

In the spontaneously broken gauge theory we have computed tree-level amplitudes for production of high multiplicities of longitudinal vector bosons and
Higgs bosons at threshold. We found that these amplitudes grow factorially $\sim m! \, n!$ with the numbers of vector and Higgs bosons in the final state, thus extending the known
factorial growth in scalar QFT to the massive Gauge-Higgs theory. We have also shown that in addition to the $m! \, n!$ factorial growth, the amplitudes involving
high multiplicities of longitudinal vector bosons grow faster than the amplitudes with only the multiple Higgs production.

These findings imply that broken gauge theory shows a very different perturbative behaviour at high multiplicities from massless gauge theories, such as QCD.
In the latter case the factorial growth of the number of Feynman diagrams is not inherited by multi-gluon amplitudes, as can be seen from e.g. the applications
of MHV rules \cite{CSW} or the BCFW recursion relations \cite{BCFW}.

\medskip
\noindent Before we conclude we would like to comment on the limitations, extensions and implications of these results.
\medskip

{\bf 1.} In this paper we have not considered the production of transversely polarised massive vector bosons. This task can be approached by solving
iteratively the full Eq.~\eqref{cleq-A} including the commutator term. It would be interesting to find out whether or not the production of $m_T$ transverse vector bosons
shows any signs of growth with respect to $m_T$ and to compare this behaviour with QCD. In either case, at tree-level this would not affect the factorial growth of longitudinal vector and Higgs production 
found here.

\medskip

{\bf 2.} What about higher loop corrections to our tree-level amplitudes at threshold? A simple scaling estimate of of the size of one-loop correction to an
$n$-point amplitude in a generic theory
is ${\cal A}_n^{\rm 1-loop} \sim \alpha \, n^2 {\cal A}_n^{\rm tree}$, where the factor of $n^2$ comes from the number of ways one can attach 
an internal propagator between $n$ external legs, and $\alpha$ represents the generic coupling constant ($\alpha_W$ or $\lambda$).
In scalar field theory this scaling argument
was confirmed in \cite{Voloshin83,Smith84} by computing the 1-loop tadpole graph in the background of the classical generating function
\eqref{noSSB} or \eqref{sol-SSB}. For example, the 1-loop corrected threshold amplitude in the real scalar QFT \eqref{eq:LSSB} is given by \cite{Smith84}
\[
\phi^4\,{\rm theory\, Eq.~\eqref{eq:LSSB}}: \quad
{\cal A}_{1\to n}^{\rm tree + 1 loop}\,=\, n!\, (2v)^{1-n} \left(1+n(n-1)\frac{\sqrt{3} \lambda}{8\pi}\right)
\,.
\label{eq:amplnh1}
\]
The ($n^2 \times\,$coupling constant) behaviour of the loop corrections does not eliminate the factorial growth found at tree-level, in fact 
at the interesting for us multiplicities, $n\sim 1/\lambda \sim 1/\alpha_W$, it indicates
a complete breakdown of the fixed order perturbation theory for high multiplicity amplitudes at threshold. Based on the rather general nature of the scaling argument,
we expect that this conclusion also applies to the massive Gauge-Higgs theory.

There are strong indications, based on the analysis of leading singularities of the multi-loop expansion around singular generating functions in scalar field theory,
that the 1-loop correction exponentiates \cite{LRST},
\[
{\cal A}_{1\to n}\,=\,{\cal A}_{1\to n}^{\rm tree}\times\, \exp\left[B\, \lambda n^2\,+\, {\cal O}(\lambda n)\right]\,
\label{expB} 
\]
in the limit $\lambda \to 0,$ $n\to \infty$ with $\lambda n^2$ fixed. Here $B$ is the constant factor determined from the 1-loop calculation,
\begin{eqnarray}
\phi^4\,{\rm theory\, Eq.~\eqref{eq:Lphi}}: \quad
B &=& -\, \frac{1}{64 \pi^2}\left(\log(7+4\sqrt{3})-i\pi\right) \,, \label{BnoSSB} \\
\phi^4\,{\rm theory\, Eq.~\eqref{eq:LSSB}}: \quad
B &=& +\, \frac{\sqrt{3}}{8 \pi} \,, \label{BSSB}
\end{eqnarray}
where the last equation is in agreement with \eqref{eq:amplnh1} and leads to the exponential enhancement of the tree-level threshold amplitude
at least in the leading order in $n^2 \lambda$. It would be interesting to investigate how the vector boson emission and loops would affect  this result and if the
overall sign of ${\rm Re}(B)$ in the Gauge-Higgs theory will remain positive. Of course, the higher-order corrections $\sim n\lambda$ are also important, as we are 
interested in multiplicities $n \sim 1/\lambda$.

\medskip

{\bf 3.} Going near or off the threshold. 
Remarkable progress has been made in the mid 90's in understanding the scalar QFT case, see \cite{LRT} for a review of these developments. To characterise the behaviour of the amplitude off the
threshold it is convenient to define the average kinetic energy per particle (per mass) in the final state \cite{LRST}, which in the non-relativistic limit near the threshold becomes,
\[
\varepsilon \,:=\, \frac{E-nM}{nM}\, \rightarrow\, \frac{1}{2nM^2}\, \sum_{j=1}^n \vec{p}_j^{\,\,2} \,.
\label{epsdef}
\]
In the non-relativistic multi-particle limit  $n\to \infty,$ $\varepsilon \to 0$ with $\varepsilon n$ fixed, one can solve the recursion relations 
for the scattering amplitudes, but now incorporating the dependence on $\varepsilon$. For the simple $\phi^4$ theory \eqref{eq:Lphi}, the result is 
\cite{LRST} ({\it cf.}
\eqref{eq:ampln2}),
\[
{\cal A}_{1\to n}\,=\, 
n!\, \left(\frac{\lambda}{8M^2}\right)^{\frac{n-1}{2}}  \times\,\exp\left[-\frac{5}{6}\,n\varepsilon\right]
\,,
\label{eq:ampln3}
\]
which solves the recursion relations to order $\varepsilon$.

Combining these results for the off-threshold non-relativistic amplitudes \eqref{eq:ampln3} and the exponentiation of the 1-loop corrections \eqref{expB}, 
the authors of \cite{LRT} have argued that the multi-particle cross-sections in the case of scalar field theory take the form,
\[\sigma_n \, \sim\, \exp\left[{\lambda}^{-1}\, F(\lambda n, \epsilon) \right]
\,.
\]
At tree-level the `holy-grail' function $F$ in the model  \eqref{eq:Lphi} takes the form \cite{LRST,LRT},
\[
{\lambda}^{-1} F^{\rm tree} \,=\,n \left[ \log\left(\frac{\lambda n}{16}\right) - 1\right]
+ n\, \frac{3}{2}\left[ \log\left(\frac{\varepsilon}{3\pi}\right) +1\right]
- n\,\varepsilon \, \frac{17}{12}\,.
\label{LRSTF}
\]

What is particularly important for our purposes of understanding the role played by the amplitudes on threshold in the more general theory, such as the 
Gauge-Higgs theory considered in this paper, is that all terms on the {\it r.h.s.} of \eqref{LRSTF} have a clear physical interpretation in terms 
of the original tree-level threshold amplitude.
The first term in square brackets is the logarithm of the tree-level squared amplitude at threshold, $|{\cal A}_{1\to n}^{\rm tree}|^2$.
The second term is the result of integrating this constant amplitude over the non-relativistic $n$-particle phase space. Finally, the third term
on the {\it r.h.s.} of \eqref{LRSTF} is the correction coming from integrating the $\varepsilon$-dependent non-relativistic expression \eqref{eq:ampln3}.

It is then natural to conjecture that the in our Gauge-Higgs theory the high-multiplicity tree-level cross-section (e.g. with the virtual Higgs in the intermediate
state) take the form,
\begin{eqnarray}
\log \sigma_{n+m}^{\rm tree} &\sim& 2\log (d(n,m)) \,+\,  n\left[ \log\left(\frac{\lambda n}{4}\right) - 1\right]\,+\, 
m \left[ \log\left(\frac{g^2 m}{32}\right) - 1\right]  \nonumber\\
&+& n\, \frac{3}{2}\left[ \log\left(\frac{\varepsilon_h}{3\pi}\right) +1\right] \,+\, m\, \frac{3}{2}\left[ \log\left(\frac{\varepsilon_V}{\pi}\right) +1\right]
+ {\cal O}( n\,\varepsilon_h + m\, \varepsilon_V)
\,,
\end{eqnarray}
where we note that $\log\left(\frac{g^2m}{32}\right)=\log\left(\kappa^2\frac{\lambda m}{4}\right)$
and
$\varepsilon_h$ and $\varepsilon_V$ are the non-relativistic kinetic energies of the $n$ Higgs bosons and $m$ longitudinal vector bosons in the final state,
\[
\varepsilon_h \,=\, \frac{1}{2n\,M_h^2}\, \sum_{j=1}^n \vec{p}_j^{\,\,2} \,\, , \quad
\varepsilon_V \,=\, \frac{1}{2m\,M_V^2}\, \sum_{k=1}^m \vec{p}_k^{\,\,2}
\,.
\]

One can also consider computing the off-shell Gauge-Higgs amplitudes directly.
The recent progress in scattering amplitudes calculations in gauge theory based on on-shell methods is largely reliant on massless states.
 These methods were extended to incorporate one or few massive stats: the Higgs boson in \cite{DKG}, the massive vector boson
currents in \cite{BernZWcurr} and few massive particles in the BCFW rules in \cite{BKGSvrc}. Nevertheless,
these results do not capture directly the high-multiplicity production of massive states we are interested in in the broken gauge theory.

One promising way forward would be to use the approach developed in \cite{DMelnikovC} for computing colour-ordered amplitudes 
in the broken gauge theory. Combining colour-ordering with the Berends-Giele-type recursion relations, the authors of \cite{DMelnikovC} have computed 
numerically tree-level scattering amplitudes of up to 9 massive vector bosons (with generic polarisations, in a generic kinematics, and no Higgses).
It should be possible to extend these results to $\sim 20-30$ gauge and Higgs bosons with external momenta near (and also away from) the threshold.

\medskip

For the case of the scalar theory with SSB \eqref{sol-SSB} it was advocated in \cite{GorskyV} that 
the $1\to n$ process should proceed via producing a non-perturbative bubble in the intermediate state, $1\to n=1\to B \to n$,
 with the bubble $B$ developing an exponentially damping form-factor as soon as the 3-momenta of external states exceed the inverse radius of the bubble.
 The bubble interpretation was based
 on the fact that in Euclidean time the generating function \eqref{sol-SSB}
 is the kink solution of the model. The conclusion of  \cite{GorskyV} was that the bubble form-factor suppression would lead to a non-perturbative exponential 
 suppression of the $1\to n$ processes in this theory. 
 In the Gauge-Higgs theory the generating function does not have a kink form as soon as the vector boson fields are taken into account. The imaginary-time solution
 to \eqref{cleq-A3} is more of the free-fall type -- not unlike the solution \eqref{noSSB} which reaches infinite field values in finite time. This then also affects the
 Higgs field in \eqref{cleq-h3}. We do not expect that in the Gauge-Higgs theory the high-multiplicity process would proceed though a non-singlular 
 and relatively long-lived  semi-classical  state (which the bubble was in the scalar SSB case) to provide for a sharp semi-classical form-factor which would
 suppress processes of the threshold.

\medskip

{\bf 4.} The main conclusion we want to draw from the computations presented in this paper is that the energies which should be available at the next generation of 
$pp$ colliders, will kinematically allow for sufficiently high multiplicities of Higgses and longitudinal $W$'s and $Z$'s production where the electroweak sector of the Standard Model
becomes strong and the perturbation theory breaks down. To answer the question whether these very high multiplicity processes become observable and even 
unsuppressed 
requires physics beyond perturbation theory in the weak sector. This is an interesting point on its own right, as it is often generally assumed that collider phenomenology 
of the electroweak SM sector is always perturbative.

\medskip

{\bf 5.}  Finally, as already mentioned, this `perturbative' high-multiplicity $n \sim 1/\alpha_W$ production in the topologically trivial sector is a logical counterpart 
of the more complicated instanton-induced non-perturbative $B-L$ processes, which also require $n \sim 1/\alpha_W$ electroweak quanta in final states. 
There is a degree of complementarity between these two processes, as one would expect that if the non-perturbative $B-L$ processes passing over the
sphaleron barrier become observable, 
the complementary perturbative processes at same energies and multiplicities should become strong (or non-perturbative).
Also it is worthwhile pointing out that the
phenomenological collider signatures of the final state for these two types of processes would be largely indistinguishable unless one measures the net charge of
 electrically charged leptons in the final state. If it will turn out that the very high multiplicity processes in the topologically trivial sector become observable, 
 there will be fewer theoretical obstacles for the B-L processes overcoming their exponential suppression in high energy collisions.

\bigskip

\section*{Acknowledgements}

I am grateful to Jeppe Andersen, Nima Arkani-Hamed, Tao Han, Michelangelo Mangano and Andreas Ringwald for useful discussions 
of perturbative and non-perturbative aspects of electroweak dynamics at  $\sim$100 TeV.
This work started at the `BSM prospects for Future Circular Colliders' meeting at CERN and the CFHEP Symposium in Beijing.
The research is supported by STFC, the Wolfson Foundation and Royal Society.
%

%
%
 
\bigskip

\bibliographystyle{h-physrev5}

\end{document}